\documentclass[%
reprint,
superscriptaddress, 
amsmath,amssymb,aps,]{revtex4-2}
\usepackage{graphicx}
\usepackage{dcolumn}
\usepackage{bm}
\usepackage[noline]{algorithm2e}
\usepackage{newtxmath}
\usepackage{amsmath}
\usepackage{graphicx}
\usepackage{tikz}
\usepackage{braket}
\usetikzlibrary{matrix,calc}

\newcommand{\mymatrix}[1]{\ensuremath{\left\downarrow\vphantom{#1}\right.\overset{\xrightarrow[\hphantom{#1}]{\text{\normalsize time}}}{#1}}}
\makeatletter
\newcommand*{\rom}[1]{\expandafter\@slowromancap\romannumeral #1@}

\begin{document}
\title{Sparse inference and active learning of stochastic differential equations from data}

\author{Yunfei Huang}
\affiliation{Theoretical Physics of Living Matter, Institute of Biological Information Processing and Institute for Advanced Simulation, Forschungszentrum Juelich, 52425 Juelich, Germany}
\affiliation{Contributed equally}
\author{Youssef Mabrouk}
\affiliation{Institute for Infectious Diseases and Zoonoses, Department of Veterinary Sciences, Ludwig-Maximilians-Universitaet Munich, 80539 Munich, Germany}
\affiliation{Helmholtz Institute Muenster (HI MS), IEK-12 Forschungszentrum Juelich GmbH, Corrensstraße 46, 48149 Muenster, Germany}
\affiliation{Contributed equally}
\author{Gerhard Gompper}
\affiliation{Theoretical Physics of Living Matter, Institute of Biological Information Processing and Institute for Advanced Simulation, Forschungszentrum Juelich, 52425 Juelich, Germany}

\author{Benedikt Sabass}
\affiliation{Theoretical Physics of Living Matter, Institute of Biological Information Processing and Institute for Advanced Simulation, Forschungszentrum Juelich, 52425 Juelich, Germany}
\affiliation{Institute for Infectious Diseases and Zoonoses, Department of Veterinary Sciences, Ludwig-Maximilians-Universitaet Munich, 80539 Munich, Germany}
\affiliation{E-mail address: b.sabass@fz-juelich.de}	
\date{\today}

\begin{abstract}
Automatic machine learning of empirical models from experimental data has recently become possible as a result of increased availability of computational power and dedicated algorithms. Despite the successes of non-parametric inference and neural-network-based inference for empirical modelling, a physical interpretation of the results often remains challenging. Here, we focus on direct inference of governing differential equations from data, which can be formulated as a linear inverse problem. A Bayesian framework with a Laplacian prior distribution is employed for finding sparse solutions efficiently. The superior accuracy and robustness of the method is demonstrated for various cases, including ordinary, partial, and stochastic differential equations. Furthermore, we develop an active learning procedure for the automated discovery of stochastic differential equations. In this procedure, learning of the unknown dynamical equations is coupled to the application of perturbations to the measured system in a feedback loop. We show that active learning can significantly improve the inference of global models for systems with multiple energetic minima.
\end{abstract}

\maketitle

\section{Introduction}
Throughout the natural sciences, mathematical models are frequently formulated as differential equations. For example, with stochastic, ordinary, and partial differential equations (SDEs, ODEs, and PDEs). In physics, governing differential equations are often derived from first principles, for instance, from conservation of energy, momentum, and thermodynamic considerations. However, for complex systems studied, e.g., in biophysics, climate science, and neuroscience, first principles determining the system properties are typically not fully known. For example, because such systems are in a driven non-equilibrium state, highly nonlinear, and because dynamics may occur on multiple scales that are not well separated. In these cases, one can resort to phenomenological, effective descriptions that may result from some level of coarse graining and are based on experimental data. Recently, the increased availability of computational power has made it possible to construct such models in an automated fashion, which is known as data-driven discovery of governing equations.

Various approaches have been developed for inferring the differential equations that govern a non-linear dynamical system directly from measured data~\cite{schmidt2009distilling,bongard2007automated,brunton2016discovering,rudy2017data,boninsegna2018sparse, raissi2020hidden}. In a popular approach called ``symbolic regression'', function libraries are employed to automatically extract the terms in a governing equation that best represents the measured data according to some optimization criterion~\cite{schmidt2009distilling,bongard2007automated}. Recently, the use of sparse regression techniques for symbolic regression has received considerable scientific attention~\cite{brunton2016discovering,rudy2017data}. In symbolic regression, the physical quantity $z$, which is for illustration taken to be a scalar here, is assumed to obey an equation of the general form
\begin{equation}
\check{z}=\mathcal{F}(z,\,\mathbf{x},\,t,\,c),    
\label{eq_generic_form}
\end{equation}
where $\check{z}$ can be, e.g., a time derivative $\check{z}=\frac{\partial z}{\partial t}$ for ODEs and PDEs. $\mathcal{F}(z,\,\mathbf{x},\,t,\,c)$ is an unknown function whose arguments $\mathbf{x}$ represent space coordinates while $t$ represents time and $c$ is a constant parameter. The aim of symbolic regression is to estimate the function $\mathcal{F}(\ldots)$ from a data set $\mathbf{z}$, which could be a measured sequence of values of $z$ at different time-space coordinates. The vector $\check{\mathbf {z}}$ is either measured or estimated from $\mathbf {z}$, e.g., with a discrete difference scheme. For inference of $\mathcal{F}(\cdots)$, a so-called ``library'' $\mathbf{\Theta(z)}$ is constructed from a suitable set of functions of $\mathbf{z}$, e.g., various powers of $\mathbf{z}$, combinations of partial derivatives, or trigonometric functions. Assuming that the governing equation (\ref{eq_generic_form}) can be expressed as a linear superposition of library terms, we write
\begin{equation}
\check{\mathbf {z}}=\mathbf{\Theta(z)}\boldsymbol{\xi},
\label{eq:1}
\end{equation}
where $\boldsymbol{\xi}$ is a weight vector. The inference of the governing equation is thus reduced to a regression problem for the optimal $\boldsymbol{\xi}$, given $\check{\mathbf {z}}$ and $\mathbf{\Theta(z)}$. In general, solving the inverse problem in Eq.~(\ref{eq:1}) is not straight-forward since the matrix $\mathbf{\Theta}$ should represent many equation terms and can have a large condition number $\kappa(\mathbf{\Theta})$.

In Ref.~\cite{brunton2016discovering}, a method called sparse identification of nonlinear dynamics (SINDy) has been proposed. The method works iteratively. At each iteration, $\boldsymbol{\xi}$ is first obtained from a least-squares optimization involving Eq.~(\ref{eq:1}) and $\boldsymbol{\xi}$ is subsequently  thresholded such that values smaller than a cutoff $\varkappa$ are set to zero. The iteration is continued until convergence conditions are satisfied. SINDy has been shown to be a powerful and versatile method that is applicable for inference of various types of ODEs~\cite{brunton2016discovering}. However, the method requires the user to manually select the thresholds $\varkappa$. For the identification of PDEs, an alternative algorithm called train sequential threshold ridge regression (TrainSTRidge) has been described in Ref.~\cite{rudy2017data}. This method is a variant of a least-squares optimization procedure for ridge regression called Sequential Threshold Ridge regression (STRidge). In STRidge, the vector $\boldsymbol{\xi}$ is first calculated by using ridge regression with a fixed regularization parameter. Then, all elements in $\boldsymbol{\xi}$ that have a smaller absolute value than a threshold $\varkappa$ are set to zero. Both, the regularization parameter and the threshold $\varkappa$ need to be provided by the user in STRidge. TrainSTRidge~\cite{rudy2017data} employs L0 regularization and a training step to automatically determine the threshold $\varkappa$ while the regularization parameter remains to be set by the expert user. Conversely, a method called threshold sparse Bayesian regression, which also was employed for identification of PDEs~\cite{zhang2018robust}, requires no input of a regularization parameters but some thresholds remain to be provided by the user.

The first aim of this work is to provide a method to solve the inverse problem associated with data-driven discovery of governing physical equations, Eq.~(\ref{eq:1}), by combining a Bayesian approach with a automatic thresholding procedure. We call this method \textit{automatic threshold sparse Bayesian learning} (ATSBL). Our algorithm does not require any manual fine-tuning of parameters to correctly infer governing differential equations from measured data. The method can be employed to identify ODEs, PDEs, and SDEs.

The case of SDEs requires particular attention, since the above-mentioned methods of equation inference are mainly designed for deterministic processes and some moderate amount of additive noise. The question {of} how to reconstruct the force fields for stochastic processes has been investigated in numerous studies, e.g., for application in soft matter physics and biophysics~\cite{bishwal2007parameter, friedrich2011approaching, stephens2011emergence, sarfati2017maximum, perez2018high, boninsegna2018sparse, baldovin2019langevin}. Recently, sophisticated methods have been proposed for dealing with discretization and the inference problem in the context of SDEs for second-order dynamics~\cite{frishman2020learning, ferretti2020building,bruckner2020inferring, bruckner2021learning}. Here, we focus on the use of symbolic regression for the inference of analytical expressions of SDEs of the overdamped Langevin-type. One approach to symbolic regression in this context is based on dividing the phase space into small hypercubes which are also called bins in the one-dimensional case. Average values of the state variables and of their derivatives are estimated in each hypercube and the regression is defined with respect to these averages~\cite{boninsegna2018sparse}. This kind of averaging generally depends on the chosen discretization and the averaging may lead to a substantial loss of information. Furthermore, application of this method to non-stationary processes requires a large ensemble of trajectories and considerable numerical effort to sample the time-dependent probability distribution in phase space. The difficulties related to the averaging in phase space motivate the investigation of the question to what extent the above-mentioned inference tools can be used in the context of noisy data without the need to perform \textit{ad hoc} averaging, and, eventually, how the robustness of the inference methods may be improved in this context. We show that imposing Laplacian or Gaussian prior distributions on the inferred models is generally sufficient to identify the correct SDEs directly from trajectories without phase-space binning and we provide a comparison of the accuracy of results obtained with the two types of prior distributions. A remarkable performance of the Laplacian prior is demonstrated with several examples, including Brownian motion in time-dependent potentials.

A major challenge for the inference of SDEs is that the phase space is often sampled very inhomogeneously in available data. This problem is encountered, e.g., for systems where the long-term dynamics is dominated by transitions between different, locally stable states, while the short-term dynamics are dominated by fluctuations around individual stable states. In such cases, the inferred equation may be meaningful only locally, i.e, within the region covered by the measurement trajectory, and it may be \textit{a priori} impossible to infer the global dynamics from a given data set. To enable an automatic inference of a global model under these conditions, we consider the question of how to design an external perturbation to the system, also called ``control force'', such that the state variables are forced to explore the full phase space in a shortened sampling time. Established Umbrella sampling routines used for this purpose rely on quadratic control forces and involve non-trivial design steps for the control force~\cite{torrie1977nonphysical, umbrella_1, umbrella_2, umbrella_3}. See, e.g., Refs.~\cite{besold1999efficient,goedecker2004minima} for alternative approaches. This kind of methodology has proved useful, e.g., in the context of computational studies of nucleation~\citep{nucleation} and growth~\citep{growth} processes. We develop an alternative adaptive control technique that recursively infers the governing equation and adapts the external control solely based on inferred equations. The adaptation loop consists of inference of the governing equation and a subsequent update of the control force such that it is directly opposite to the inferred force. No parameters need to be tuned for designing the control with this adaptive scheme. Using the adaptive control scheme, we demonstrate a substantial improvement of the inference of SDEs for several different simulations of Brownian motion.  

This work is organized as follows. The Methods section provides details on the the construction of function libaries and the casting of the inference problem into a system of linear equations. The inference algorithm is summarized and it is explained how Laplacian prior distributions can be used to impose the sparsity condition on the inferred models. In the Results section, the performance of the described method is illustrated by means of numerical examples and a comparison with previously described methods is presented. An adaptive sampling technique for improving the inference of SDEs is proposed and the usefulness of this approach is demonstrated. 

\section{Methods for data-driven identification of differential equations}

\subsection{Ordinary and partial differential equations}
Measurement data from a system of interest is presumed to be recorded as a time series of states, for example, a time-dependent position vector. In a data-driven approach to model a system, the data is used to automatically infer the \textit{a priori} unknown dynamical equations that govern the observed process. In this work, inference is based on libraries of candidate functions for the governing equations. The data used for inference of differential equations is assumed to contain additive noise but no systematic errors.

For inference of ODEs, we generalize the introductory example for a scalar variable $z$, Eq.~(\ref{eq:1}), to a system with $M$ components that are assumed to be sampled with the same regular time interval for all $\ell \in \{1\ldots M\}$ components. {To distinguish discrete measurements from continuous variables, a subscript notation is employed in the following.} The $\ell$-th component measured in an ordered time series $[t_1,\dots,t_N]$ is written as $\mathbf{z}_{\ell} =[z_{\ell,t_1},\;z_{\ell,t_2},\;\dots,\;z_{\ell,t_N}]$. {Vectors or arrays containing multiple variable measurements, e.g., at different time points, are denoted with bold letters.} The whole data can then be written in matrix form as
\begin{equation*}
\mathbf{Z}=[\mathbf{z}_1^{\mathrm{T}},\;\mathbf{z}_2^{\mathrm{T}},\;\dots,\;\mathbf{z}^{\mathrm{T}}_M] = 
\rotatebox[origin=c]{90}{\text{state}}\mymatrix{\begin{bmatrix}
	z_{1, t_1} & z_{1, t_2} & \cdots & z_{1, t_N} \\
	z_{2, t_1} & z_{2, t_2} & \cdots & z_{2, t_N} \\
	\vdots   & \vdots   & \ddots & \vdots  \\
	z_{M, t_1} & z_{M, t_2} & \cdots & z_{M, t_N}
	\end{bmatrix}}.
\end{equation*}
{Our approach also requires derivatives of the measured data. For simplicity, finite-difference approximations are used throughout this work. Approximate derivatives are denoted by the operator $\mathcal{D}_{\cdots}$, which represents here a fourth-order finite central difference scheme. For example, a time derivative of the $\ell$-th state component, $\mathbf{z}_{\ell}$, at the $i$-th timepoint $t_i$ is written as ${\dot{z}_{\ell}(t)}|_{t=t_i} \approx \mathcal{D}_t\mathbf{z}_{\ell}|_{t=t_i}$. For the entire dataset, we write the time derivative as}
\begin{equation*}
\dot{\mathbf{Z}} \approx \mathcal{D}_t \mathbf{Z}= [\mathcal{D}_t\mathbf{z}^{\mathrm{T}}_1,\;\mathcal{D}_t\mathbf{z}^{\mathrm{T}}_2,\;\dots,\;\mathcal{D}_t\mathbf{z}^{\mathrm{T}}_M].
\end{equation*}
A governing ODE for the vector containing the trajectory of the $\ell$-th state component may be written as a linear combination of elementary functions of all $\{\mathbf{z}_{\ell'}\}$, e.g., {as
\begin{equation}
\mathcal{D}_t{\mathbf{z}}_{\ell}= \mathcal{F}_{\ell}(\{\mathbf{z}_{\ell'}\},\;\{\mathbf{z}_{\ell''}\odot \mathbf{z}_{\ell'''}\},\;\ldots,\;\{\cos \mathbf{z}_{\ell'}\},\;\ldots,c),
\label{eq:5-2} 
\end{equation}
where the indices $\ell'$, $\ell''$, and $\ell'''$ cover the $M$ system dimensions,
$\odot$ denotes an element-wise product,} and $c$ represents a constant. {$\mathcal{F}_{\ell}$ can also depend on time, but we focus mostly on autonomous differential equations in the following.} Since $\mathcal{F}_{\ell}(\cdot)$ represents a linear combination of functions that can be calculated from the data, $\mathcal{F}_{\ell}(\cdot)$ can be expressed with the help of a library matrix $\mathbf{\Theta}(\mathbf{Z})$ multiplied with a sparse vector $\boldsymbol{\xi}_{\ell}$. Thus, {we obtain for Eq.~(\ref{eq:5-2}) in discretized form} 
\begin{equation}
\mathcal{D}_t\mathbf{z}_{\ell}=\mathbf{\Theta}(\mathbf{Z})\boldsymbol{\xi}_{\ell},
\label{eq:5-40}
\end{equation}  
where the terms of the library matrix $\mathbf{\Theta}(\mathbf{Z})$ are calculated from the measurement data by evaluating the functions of $\{\mathbf{z}_{\ell'}\}$ {and the non-zero elements of $\boldsymbol{\xi}_{\ell}$ characterize the dynamics of the system}. Since Eq.~(\ref{eq:5-40}) refers to ODEs, no derivative terms are contained in the library on the right-hand side of the equation. Given $\mathcal{D}_t{\mathbf{z}}_{\ell}$ and $\mathbf{\Theta}(\mathbf{z})$, the aim is to calculate a sparse vector $\boldsymbol{\xi}_{\ell}$ with a minimal number of non-zero coefficients corresponding to a minimal number of terms necessary to describe the dynamics.

For inference of PDEs, the library matrix $\mathbf{\Theta}$ has to contain partial-derivative terms. Thus, data is required that allows the numerical estimation of derivative expressions with respect to two or more variables, for example, with respect to time and space. Usually, measurements therefore consist of discrete space-time series recordings of system variables. For example, an array $\mathbf{Z}^P$ representing the $M$-dimensional state vector that is measured at $N$ time points in $R$ positions of one space coordinate $x$ is written as 
\begin{equation*}
\begin{tikzpicture}[every node/.style={anchor=north east,fill=white,minimum width=0.2cm,minimum height=1mm}]
\matrix (mA) [draw,matrix of math nodes]
{
	z_{1,(t_1,x_R)} & z_{1,(t_2,x_R)} & \cdots & z_{1,(t_N,x_R)} \\
    z_{2,(t_1,x_R)} & z_{2,(t_2,x_R)} & \cdots & z_{2,(t_N,x_R)} \\
    \vdots   & \vdots   & \ddots & \vdots  \\
    z_{M,(t_1,x_R)} & z_{M,(t_2,x_R)} & \cdots & z_{M,(t_N,x_R)}\\
};

\matrix (mB) [draw,matrix of math nodes] at ($(mA.south west)+(4.89,2)$)
{
	z_{1,(t_1,x_2)} & z_{1,(t_2,x_2)} & \cdots & z_{1,(t_N,x_2)} \\
    z_{2,(t_1,x_2)} & z_{2,(t_2,x_2)} & \cdots & z_{2,(t_N,x_2)} \\
    \vdots   & \vdots   & \ddots & \vdots  \\
    z_{M,(t_1,x_2)} & z_{M,(t_2,x_2)} & \cdots & z_{M,(t_N,x_2)}\\
};

\matrix (mC) [draw,matrix of math nodes] at ($(mB.south west)+(4.89,2)$)
{
	z_{1,(t_1,x_1)} & z_{1,(t_2,x_1)} & \cdots & z_{1,(t_N,x_1)} \\
    z_{2,(t_1,x_1)} & z_{2,(t_2,x_1)} & \cdots & z_{2,(t_N,x_1)} \\
        \vdots   & \vdots   & \ddots & \vdots  \\
    z_{M,(t_1,x_1)} & z_{M,(t_2,x_1)} & \cdots & z_{M,(t_N,x_1)}\\
};
\draw[dashed](mA.north east)--(mC.north east);
\draw[dashed, anchor=center](mA.north west)-- node[sloped,above] {space} (mC.north west);
\draw[dashed](mA.south east)--(mC.south east);

\node [above left, anchor=center] at (mC.south) {time};
\node [above left, rotate=90, anchor=mid] at (mC.west) {state};
\node [above left, xshift=-0.1cm, yshift=-0.25cm] at (mC.west) {$\mathlarger{\mathlarger{\mathlarger{\mathbf{Z}^P}}}=$};
\end{tikzpicture}
\end{equation*}
With a finite-difference approximation, vectors of time derivatives of every component, {$\mathcal{D}_t\mathbf{z}_{\ell}$, and various orders of $x$ derivatives are calculated, for example, $\mathcal{D}_x\mathbf{Z}^P,\;\mathcal{D}_{xx}\mathbf{Z}^P,\;\dots$}. These derivative terms are added to the library $\mathbf{\Theta}^P$. Like for ODEs, inference of the dynamical equation governing the component $\mathbf{z}_{\ell}$ is then based on the linear equation
\begin{equation}
\mathcal{D}_t{\mathbf{z}}_{\ell}=\mathbf{\Theta}^P\left(\mathbf{Z}^P,\mathcal{D}_{x}\mathbf{Z}^P,\mathcal{D}_{xx}\mathbf{Z}^P,\ldots\right)\boldsymbol{\xi}_{\ell},
\label{eq:5-41}%
\end{equation} 
with a sparse coefficient vector $\boldsymbol{\xi}_{\ell}$ to be determined. 

{Note that a robust estimation of derivatives from noisy data is an important prerequisite for data-driven inference of ODEs and PDEs in this framework. The fourth order finite-difference approximations employed here may be supplemented or replaced with other methods, including denoising procedures and Gaussian process regression models.}

\subsection{Stochastic differential equations}
We focus on Langevin-type SDEs to describe the time evolution of continuous, real state variables $\mathbf{X}(t)$, representing, e.g., the position of a Brownian particle in space~\cite{risken1996fokker}. Trajectories, denoted by $\mathbf{X}(t)$, are time-ordered sequences of values of space coordinates $\mathbf{x}$. The general form of the considered SDEs is
\begin{equation}
\mathrm{d}X_{\ell}(t)= \underbrace{g_{\ell}(\mathbf{X}(t),t)\mathrm{d}t}_\text{deterministic part}+\underbrace{h_{\ell,\ell'}(\mathbf{X}(t),t)\, \mathrm{d} W_{\ell'}(t)}_\text{noise}, 
\label{eq:2}
\end{equation} 
where we employ the Einstein sum convention and $X_{\ell}(t)$ denotes the $\ell$-th component of the system state at time $t$. The trajectories $\mathbf{X}$ are calculated by making use of Ito's interpretation of stochastic integrals~\cite{risken1996fokker}. The $g_{\ell}(\mathbf{X}(t),t)$ represent the deterministic parts of the differential equations. For example, for a Brownian particle undergoing overdamped motion in the presence of conservative forces with a potential $U(\mathbf{x},t)$, we have $g_{\ell}(\mathbf{X},t)= -\nabla_{x_{\ell}} U(\mathbf{x},t)|_{\mathbf{x}=\mathbf{X}(t)}$. The {stochastic perturbations} are assumed to result from a Wiener process with a noise source $\Gamma_{\ell}(t)$ and $\mathrm{d} W_{\ell} =\Gamma_{\ell}(t)\,\mathrm{d} t$. The noise is assumed to obey a Gaussian distribution with a vanishing mean and a $\delta$-correlated variance as 
\begin{subequations}
\begin{align}
\langle \Gamma_{\ell}(t) \rangle&=0,\\
\langle \Gamma_{\ell}(t)\Gamma_{\ell'}(t') \rangle&=\delta_{\ell,\ell'}\delta(t-t'),
\end{align}
\end{subequations}
respectively. {The coefficient matrix $h_{\ell,\ell'}$ in Eq.~(\ref{eq:2}) scales the magnitude of the stochastic perturbations and is assumed to be diagonal, for simplicity. Further noise sources, e.g., resulting from an experimental measurement of a trajectory, are not explicitly considered throughout this work.}

The Fokker-Planck equation that corresponds to Eq.~(\ref{eq:2}) and describes the evolution of a probability density function $f(\mathbf{x},t)$ is given by
\begin{equation}
\frac{\partial f(\mathbf{x},t)}{\partial t} = \hat{L}f(\mathbf{x},t), 
\label{eq:5-44}
\end{equation} 
where the Fokker-Planck operator $\hat{L}$ acting on $f(\mathbf{x},t)$ has the form
\begin{equation}
\hat{L}f(\mathbf{x},t)= -\frac{\partial}{\partial x_{\ell}}D_{\ell}^{(1)}(\mathbf{x},t)f(\mathbf{x},t) +\frac{\partial^2}{\partial x_{\ell} \partial x_{\ell'}}D_{\ell,\ell'}^{(2)}(\mathbf{x},t)f(\mathbf{x},t). 
\label{eq:5-45}
\end{equation}
The functions $D_{\ell}^{(1)}(\mathbf{x},t)$ and $D_{\ell,\ell'}^{(2)}(\mathbf{x},t)$ are called Kramers-Moyal (KM) coefficients or drift and diffusion coefficients. Under the assumption of perfect knowledge of the trajectories $\mathbf{X}(t)$, the KM coefficients can be calculated from the incremental changes $\Delta X_{\ell}(t) \equiv X_{\ell}(t+\tau)-X_{\ell}(t)$ in an infinitesimal time interval $\tau$ as
\begin{subequations}
\begin{align}
&D_{\ell}^{(1)}(\mathbf{x},t) = \lim_{\tau\to 0}\frac{1}{\tau}\langle [\Delta X_{\ell}(t)] \rangle_{\mathbf{X}(t)=\mathbf{x}}, \label{eq:3} \\
&D_{\ell,\ell'}^{(2)}(\mathbf{x},t)= \lim_{\tau\to 0}\frac{1}{2\tau}\langle [\Delta X_{\ell}(t)] [\Delta X_{\ell'}(t)] \rangle_{\mathbf{X}(t)=\mathbf{x}} \label{eq:4},
\end{align}
\label{eq:3-4}\\
\end{subequations}
where {$\langle \ldots \rangle_{\mathbf{X}(t)=\mathbf{x}}$ denotes averages over the stochastic trajectories.} The KM coefficients are related to the functions $g_{\ell}$ and $h_{\ell,\ell'}$ in the Langevin equation as
\begin{subequations}
\begin{align}
&g_{\ell}(\mathbf{x},t)= D_{\ell}^{(1)}(\mathbf{x},t), \\
&h_{\ell,\ell'}(\mathbf{x},t)= \sqrt{2D_{\ell,\ell'}^{(2)}(\mathbf{x},t)}.
\end{align}
\end{subequations}
We consider only diagonal diffusion matrices, but the KM coefficients can depend explicitly on space and time. To estimate the coefficients, $M$-dimensional trajectories $X_{\ell,i}$, $\ell \in \{1,\dots,M\}$ are sampled with a small, regular time step $s$ at time points $i \in \{1,\dots,N\}$. {Trajectory samples $X_{\ell,i}$ are distinguished from the original stochastic variable $X_{\ell}(t)$ by the index $i$, representing the $i$-th time point.} Therewith, two new sequences are constructed as
\begin{subequations}
\begin{align}
&\mathbf{F}^{(1)}_{\ell} =\{F^{(1)}_{\ell,i}\}_{i=1,\dots,N}= \bigg\{\frac{X_{\ell,i+1}-X_{\ell,i}}{s}\bigg\}_{i=1,\dots,N},\\
&\mathbf{F}^{(2)}_{\ell} =\{F^{(2)}_{\ell,i}\}_{i=1,\dots,N}= \bigg\{\frac{(X_{\ell,i+1}-X_{\ell,i})^2}{2s}\bigg\}_{i=1,\dots,N},
\end{align}
\end{subequations}
where $s$ is a small time step~\cite{boninsegna2018sparse}. 
{The $\bm{F}_\ell^{(1)}$ and $\bm{F}_\ell^{(2)}$ are constructed with sample trajectories from random processes that are not differentiable. Use of these quantities for estimation of the KM coefficients in the spirit of Eq.~(\ref{eq:3-4}) makes it necessary to first sample the stochastic process extensively to then approximate the average $\langle \ldots \rangle_{\mathbf{X}(t)=\mathbf{x}}$ over different realizations of the process.}

{Note that in basing the estimation on Eq.~(\ref{eq:3-4}), we are neglecting two problems that occur for time series measured in the ``real world''. Firstly, measurement noise may render the assumption of a Markov process invalid on small scales~\cite{kleinhans2007markov}. 
Secondly, the finite sampling interval $s$ cannot be made arbitrarily small in practice and therefore the estimated KM coefficients deviate systematically from the true coefficients~\cite{ragwitz2001indispensable,friedrich2002comment}. Procedures for correcting finite-sampling-time errors are available for various stochastic processes \cite{gottschall2008definition,honisch2011estimation, rydin2021arbitrary}. While the focus of this work is on the inference problem for governing equations, finite sampling-time corrections should be employed in practical applications.} 

In the following, we employ two different methods for estimating the drift and diffusion coefficients. Firstly, a method is described in the next subsection that is based on binning of the data in phase space to produce histograms. Secondly, we compare the results obtained from data binning with results from direct estimation of the KM coefficients.

\subsubsection{Estimation of KM coefficients from binned data}
\label{sec_binning_method}
A classical method for the characterization of {stationary, Markovian time series resulting from Langevin dynamics} is based on binning of the trajectory data in space intervals~\cite{rinn2016langevin,boninsegna2018sparse,gradivsek2000analysis}. For this approach, we focus on problems with only one space dimension ($M=1$). To estimate probability distributions, the data {from multiple sample trajectories of the stochastic process} is grouped into $Q$ bins and the values in each bin are averaged as
\begin{subequations}
\begin{align}
&\{X_{i}\}_{i=1,\dots,N} \mapsto \{\bar{X}_k\}_{k=1,\dots,Q}=\bar{\mathbf{X}},\\
&\{F^{(1)}_{i}\}_{i=1,\dots,N} \mapsto \{\bar{F}^{(1)}_k\}_{k=1,\dots,Q}=\bar{\mathbf{F}}^{(1)},\\
&\{F^{(2)}_{i}\}_{i=1,\dots,N} \mapsto \{\bar{F}^{(2)}_{k}\}_{k=1,\dots,Q}=\bar{\mathbf{F}}^{(2)},
\end{align}
\end{subequations}
where $\bar{X}_{k}$, $\bar{F}^{(2)}_{k}$, and $\bar{F}^{(2)}_{k}$ are bin-wise averages. The estimated probability for finding trajectory parts in the $k$-th bin, $p_k$, is normalized as $\sum_{k=1}^{Q}p_k=1$ with $0\le p_k\le1$. {Histograms resulting from data binning directly yield the curves for the drift and diffusion coefficients}, see Refs.~\cite{rinn2016langevin,boninsegna2018sparse}. The equations for the KM coefficients, $D^{(1)}(x)$ and $D^{(2)}(x)$, are inferred by finding analytical expressions for $\bar{\mathbf{F}}^{(1,2)}$ as functions of $\bar{\mathbf{X}}$. For this purpose, a library $\mathbf{\Theta} \in \mathbb{R}^{Q\times K}$ is constructed from the binned data, where $Q$ is the number of bins and $K$ is the number of terms in the library. 
For example, $\mathbf{\Theta}(\bar{\mathbf{X}})=[\mathbf{1},\;\bar{\mathbf{X}},\;\bar{\mathbf{X}}\odot\bar{\mathbf{X}},,\;\bar{\mathbf{X}}\odot\bar{\mathbf{X}}\odot\bar{\mathbf{X}},\; \sin(\bar{\mathbf{X}}),\; \dots]$ where $\odot$ again denotes an element-wise product. If the library contains all the function expressions necessary to describe the KM coefficients analytically, the governing equations can be written as
\begin{subequations}
	\begin{align}
	&\bar{\mathbf{F}}^{(1)} = \mathbf{\Theta}(\bar{\mathbf{X}})\mathbf{W}^{(1)}, \label{eq:5a}\\
	&\bar{\mathbf{F}}^{(2)} = \mathbf{\Theta}(\bar{\mathbf{X}})\mathbf{W}^{(2)},\label{eq:5b}
	\end{align}
	\label{eq:5}%
\end{subequations}
where $\mathbf{W}^{(1)}$ and $\mathbf{W}^{(2)}$ are two sparse vectors whose non-zero entries correspond to the library terms to be included in the sought-for analytical expressions for the KM coefficients. Equation~(\ref{eq:5a}) yields $D^{(1)}(x)$ and Eq.~(\ref{eq:5b}) yields $D^{(2)}(x)$. The inverse problems of finding optimal $\mathbf{W}^{(1,2)}$ in Eq.~(\ref{eq:5}) have the same form as the problem in Eq.~(\ref{eq:1}). 

{The binning of trajectories can produce significant errors in sparsely sampled regions, both in the interior and at the boundaries of the sampled phase space.}
We propose that the identification of SDEs can be improved by removal or filtering of the bins with high uncertainty. {To substantiate this suggestion, we implement the inference procedure for unfiltered histograms and, additionally, implement a straight-forward extension that essentially consists of fixing a small probability threshold, below which all the data is discarded. While the probability threshold can can be determined in different ways,} we employ here an automatic heuristic that was originally designed for edge detection in images~\cite{el2011new}. {The procedure that is described in Ref.~\cite{el2011new} consists of dividing the data according to probability thresholds to maximize the Shannon and Tsallis entropy, respectively.} Maximization of the Shannon entropy produces thresholds that divide the data into ``foreground'' and ``background'', corresponding to signal-dominated and noise-dominated phase-space regions, respectively. The threshold value determining the ``background'' is then improved in a second step by maximizing the Tsallis entropy, {whose pseudo additivity reportedly improves the analysis of data containing long-range correlations, see also Ref.~\cite{hamza2006nonextensive}.
While we found that this method for determining a probability threshold is useful in practice, its theoretical underpinnings are to our knowledge not entirely clear. Thus, a manual selection of the probability threshold based on the results may be preferable in some cases.}

\subsubsection{Estimation of KM coefficients without data binning} 
A more direct approach for estimating the KM coefficients is based on the use of the trajectories $\mathbf{F}^{(1)}_{\ell}$ and $\mathbf{F}^{(2)}_{\ell}$ without binning or filtering. {Since we do not intend to study transient initial dynamics, we mostly employ as input data a single, long trajectory generated from the stochastic process.}
For inference of the KM coefficients from the space-time trajectories, we construct a library $\mathbf{\Theta} \in \mathbb{R}^{N\times K}$, where $N$ is the length of the trajectory and $K$ is the number of terms in the library. For example, $\mathbf{\Theta}(\{\mathbf{X}_{\ell'}\})=[\mathbf{1},\;\mathbf{X}_{1},\;\ldots,\mathbf{X}_{M},\;\mathbf{X}_{1}\odot\mathbf{X}_{2},\; \dots,\; \sin(\mathbf{X_1}),\ldots]$, where $\ell'$ covers all $M$ components of the stochastic process. {Note that the library is constructed such that $F^{1}_{\ell,i}$ and $F^{2}_{\ell,i}$ at the $i$-th time point depend only on functions involving coordinates $\{X_{\ell',i}\}_{\ell'}$ at the same time point. Thus, a velocity dependence or a history dependence of the estimators for the drift and diffusion coefficients is excluded.}
Under the assumption that the library contains all necessary terms describing the drift and diffusion coefficients, the coefficients for the $\ell$-th component of the stochastic process can be inferred with
\begin{subequations}
	\begin{align}
	&\mathbf{F}_{\ell}^{(1)} = \mathbf{\Theta}(\{\mathbf{X}_{\ell'}\})\mathbf{W}_{\ell}^{(1)}, \label{eq:6a}\\
	&\mathbf{F}^{(2)}_{\ell} = \mathbf{\Theta}(\{\mathbf{X}_{\ell'}\})\mathbf{W}^{(2)}_{\ell},\label{eq:6b}
	\end{align}
	\label{eq:6}%
\end{subequations}
where $\ell \in \{1 \ldots M\}$ and the vectors $\mathbf{W}^{(1,2)}_{\ell}$ are non-zero in those entries that correspond to the terms in the libary that are required for the analytical description of the KM coefficient. The determination of the $\mathbf{W}^{(1,2)}_{\ell}$ is again an inverse optimization problem.

\subsection{Solution of the inference problems with automatic threshold sparse Bayesian learning}
For identification of the relevant library terms as, e.g., for Eq.~(\ref{eq:6}), we propose an algorithm that we call Automatic threshold sparse Bayesian learning (ATSBL). The method consists of two main steps. First, the inverse problem is solved with an efficient algorithm called Bayesian compressive sensing using Laplace priors (BCSL)~\cite{babacan2009bayesian}. Since the library is large, the solution vector generated by the BCSL algorithm typically still contains quite a few non-vanishing but small entries. Therefore, in a second step, the negligible contributions to the resulting governing equations are removed by an automatic thresholding procedure~\cite{brunton2016discovering,boninsegna2018sparse,zhang2018robust}. These two steps of the method are detailed below.
\subsubsection{Bayesian compressive sensing using Laplace priors (BCSL)}
We consider a generic linear equation system involving a given vector $\mathbf{g}$ and matrix $\mathbf{\Phi}$ and an unknown, sparse vector $\mathbf{w}$ as
\begin{equation}
\mathbf{g} = \mathbf{\Phi}\mathbf{w} +\mathbf{s},
\label{eq:2-2}
\end{equation} 
where the vector $\mathbf{s}$ represents noise or measurement errors. {Here, $\mathbf{w}$ can be thought of as a solution vector appearing in an iterative solution procedure for Eq.~(\ref{eq:5-41}), Eq.~(\ref{eq:5}), or Eq.~(\ref{eq:6}).} Various methods can be used to calculate sparse solution vectors $\mathbf{w}$ from Eq.~(\ref{eq:2-2}). In particular research on compressive sensing, which deals with the reconstruction of sparse signals from underdetermined systems, has yielded broadly applicable, efficient methods for finding sparse solution vectors $\mathbf{w}$. Among these are Bayesian methods based on the relevance vector machine (RVM)~\cite{tipping2001sparse, ji2008bayesian}. Very sparse result vectors are obtained if a Laplace distribution is used as a prior probability distribution for $\mathbf{w}$. Here, we employ a method called Bayesian compressive sensing using Laplace priors (BCSL)~\cite{babacan2009bayesian}. Specifically, we employ a variant of BCSL that interatively calculates approximate solutions, which is very computationally efficient and yields accurate results for our type of applications.

Briefly, the mathematical basis of BCSL is as follows, see Ref.~\cite{babacan2009bayesian}. The method is based on a three-stage hierarchical model. It is assumed that the errors $\mathbf{s}$ are drawn from a zero-mean Gaussian distribution with unknown variance $1/\beta$. Therefore, the likelihood function for finding a vector $\mathbf{g}$ is given by
\begin{equation}
p(\mathbf{g}|\mathbf{w},\beta) = \Big(\frac{2\pi}{\beta}\Big)^{-\frac{N}{2}}\exp\Big\{ -\frac{\beta}{2}\|\mathbf{g}-\mathbf{\Phi w}\|^2_2\Big\}.
\label{eq:9}
\end{equation}
The unknown vector $\mathbf{w}$ is assigned a prior distribution, which represents our knowledge on the nature of this quantity. To encode sparsity, one would like to employ a Laplace prior $p(\mathbf{w}|\lambda)= \lambda/2\exp(-\lambda\sum_{i}|w_i|/2)$ with a hyperparameter $\lambda$. However, the evaluation of integrals using this choice of a Laplace prior is not readily achieved since the Laplace prior is not conjugate to the Gaussian likelihood, Eq.~(\ref{eq:9}). Therefore, an auxiliary vector of non-negative hyperparameters $\boldsymbol{\gamma}$ with the same dimension as $\mathbf{w}$ is employed to express the prior as the convolution of the two different distributions $p(\mathbf{w}|\boldsymbol{\gamma})=\Pi_{i}\left[\exp{(-w_i^2/(2 \gamma_i))}/\sqrt{2 \pi \gamma_i } \right]$ and $p(\boldsymbol{\gamma}|\lambda)=\Pi_{i}\left[\lambda \exp{(-\lambda \gamma_i/2)}/2\right]
$. These two distributions together result in a Laplace prior after marginalizing out $\boldsymbol{\gamma}$ as
\begin{equation}
p(\mathbf{w}|\lambda)=\int_0^{\infty} p(\mathbf{w}|\boldsymbol{\gamma})p(\boldsymbol{\gamma}|\lambda) \mathrm{d} \boldsymbol{\gamma}=\frac{\lambda^{N/2}}{2^N} e^{-\sqrt{\lambda}\sum_{i}|w_i|},
\label{eq:10}
\end{equation}
see  Ref.~\cite{figueiredo2003adaptive}.
Overall, the joint probability density results as
\begin{equation}
p(\mathbf{g},\mathbf{w},\boldsymbol{\gamma},\lambda,\beta)=p(\mathbf{g}| \mathbf{w},\beta) p(\mathbf{w}|\boldsymbol{\gamma})p(\boldsymbol{\gamma}|\lambda)p(\lambda)p(\beta),
\label{eq:jointp}
\end{equation}
where the parameters $\lambda$ and $\beta$ are both assumed to obey Gamma distributions.
To infer values for the most probable solution vector $\mathbf{w}$ as well as the hyperparameters, an evidence procedure is employed wherein the posterior probability $p(\mathbf{w},\boldsymbol{\gamma},\lambda,\beta|\mathbf{g})$ is maximized with respect to $\mathbf{w}$, $\boldsymbol{\gamma}$, $\lambda$, and $\beta$, given the data. By making use of the expression 
\begin{equation}
p(\mathbf{w},\boldsymbol{\gamma},\lambda,\beta|\mathbf{g})=p(\mathbf{w}|\mathbf{g},\boldsymbol{\gamma},\lambda,\beta)p(\boldsymbol{\gamma},\lambda,\beta|\mathbf{g})=\frac{p(\mathbf{g},\mathbf{w},\boldsymbol{\gamma},\lambda,\beta)}{p(\mathbf{g})}
\label{eq:bays}
\end{equation} together with Eq.~(\ref{eq:jointp}), $\mathbf{w}$ is determined by simply maximizing $p(\mathbf{g}| \mathbf{w},\beta) p(\mathbf{w}|\boldsymbol{\gamma})$. This calculation yields for the result vector the expression $\mathbf{w}^*=\beta\mathbf{\Sigma}\mathbf{\Phi^T}\mathbf{g}$ with $\mathbf{\Sigma} = (\beta\mathbf{\Phi^T\Phi}+\mathbf{\Lambda})^{-1}$ with $\boldsymbol{\Lambda} $ being a square matrix that contains the $(1/\gamma_i)$ on the diagonal and is zero otherwise. This step corresponds to a Ridge regression that depends on the unknown values of $\boldsymbol{\gamma}$, $\lambda$, and $\beta$. Determination of these hyperparameters proceeds by maximizing
\begin{equation}
p(\boldsymbol{\gamma},\lambda,\beta | \mathbf{g})=\frac{ p(\boldsymbol{\gamma},\lambda,\beta, \mathbf{g})}{p(\mathbf{g})}
\label{eq:bays2}
\end{equation}
with respect to  $\boldsymbol{\gamma}$, $\lambda$, and $\beta$. Here, $p(\boldsymbol{\gamma},\lambda,\beta, \mathbf{g})$ is calculated from the right hand side of Eq.~(\ref{eq:jointp}) by integrating out $\mathbf{w}$. With the fast, approximate version of BCSL, the equations determining the optimal values of the hyperparameters are solved iteratively, where only one entry of the vector $\boldsymbol{\gamma}$ is adjusted in every step.

\subsubsection{Automatic thresholding}
Solution of the inverse problem (\ref{eq:2-2}) with BCSL typically yields vectors $\mathbf{w}$ that contain only a few large entries, but also a number of very small, non-zero entries. Removal of these negligible entries is desirable and we improve the solution sparsity with an iterative thresholding procedure~\cite{rudy2017data}. The pseudocode~\ref{alg1} illustrates how the thresholding procedure proposed in in Ref.~\cite{rudy2017data} is combined with BCSL proposed in Ref.~\cite{babacan2009bayesian}.
Briefly, the thresholding algorithm works as follows. The input is given by $\mathbf{g}$, the library matrix $\mathbf{\Theta}$, an initial increment $d_{\text{tol}}$ for the threshold $tol$, and the maximum number of iterations $n_{\text{iters}}$. The data $\mathbf{g}$ and $\mathbf{\Theta}$ is spilt into two parts for training and test, respectively. Usually, 80\% of the data is used for training and 20\% for testing. Thresholds are calculated iteratively from the training data and the validity of the thresholds is evaluated based on the error resulting from their application to the test data.
The core part of the algorithm is a loop for iterative calculation of the sparse vector $\mathbf{w}$ and the threshold $tol$. In each iteration step,  the approximate, fast BCSL routine is first employed to obtain an estimate of $\mathbf{w}$ from the training data. 
The quality of this solution estimate is evaluated by calculating the resulting error with the test data
\begin{equation}
e = \|\mathbf{\Theta} ^{\text{test}}\mathbf{w}- \mathbf{g}^{\text{test}}\|^2_2+\eta\|\mathbf{w}\|_0,
\label{eq:5-4}
\end{equation} 
where the penalty factor of the solution norm is chosen $\eta =10^{-3}\,\kappa(\mathbf{\Theta})$ as suggested for the original algorithm~\cite{rudy2017data}. If the error of the current solution is smaller than the error of previous iterations, the new solution is accepted and the threshold $tol$ is increased. In the opposite case, the threshold is decreased and the increment $d_{\mathrm{tol}}$ is refined. The final solution $\mathbf{w}_{\text{best}}$ is the sparse vector that determines the terms in the governing differential equations, SDEs, ODEs, and PDEs.

\begin{algorithm}[hbt!]
	\SetAlgoLined
	\SetKwFor{For}{for (}{) $\lbrace$}{$\rbrace$}
	\SetKwInput{KwFunction}{Function}
	\KwFunction{ATSBL($\mathbf{\Theta}$, $\mathbf{g}$, $d_{\mathrm{tol}}$, $n_{\mathrm{iters}}$)}
	\%\% Split data into parts for training and test \\
	$\mathbf{\Theta} \mapsto [\mathbf{\Theta} ^{\text{train}}, \mathbf{\Theta} ^{\text{test}}]$ \space\space\space \% ca. 80\% training, 20\% test \\
	$\mathbf{g} \mapsto [\mathbf{g}^{\text{train}}, \mathbf{g}^{\text{test}}]$\\
	$\,$\\
	\%\% Initialization\\
	$\sigma^2 = \mathrm{var}(\mathbf{g}^{\text{train}})$  \space\space \% Variance of data.\\
	$\eta =10^{-3}\kappa(\Theta)$ \space\space\space\space\space\space \% $\kappa(\Theta)$ is condition number\\
	$tol = d_{\text{tol}}$ \space\space\space\space\space\space\space\space\space\space\space \% Initial threshold \\ 
	$Q =$ \text{size} $(\mathbf{\Theta},2)$ \space\space\space\space \% Number of library terms \\
	$\mathbf{w}_{\text{best}} = $ ($\mathbf{\Theta} ^{\text{train}})^{-1}\mathbf{g}^{\text{train}}$\space\space\space\space\space\space\space\space \%Initial solution guess\\
	$e_{\text{best}} = \|\mathbf{\Theta} ^{\text{test}}\mathbf{w}_{\text{best}}-\mathbf{g}^{\text{test}}\|^2_2+\eta\|\mathbf{w}_{\text{best}}\|_0$ \%Initial error\\
	$\,$ \\
	\%\% Solution with iterative threshold adaptation\\
	\For{$i = 1;\ i < n_{\mathrm{iters}};\ i = i + 1$}{
	    $\mathbf{\hat{w}} =$ FastBCSL$(\mathbf{\Theta} ^{\text{train}},\mathbf{g}^{\text{train}}, \sigma^2)$\\
	    bigcoeffs =\{$m: |\hat{w}_m| \geq tol$\}\\
	    $\mathbf{\Theta}^{\text{train}}_{\text{old}} = \mathbf{\Theta}^{\text{train}};$\\
	    $\mathbf{\Theta}^{\text{train}} = \mathbf{\Theta}^{\text{train}}(:,$bigcoeffs$)$\\
	        \eIf{$i==1$}{$final_p = \text{bigcoeffs}$; \\
	        $\mathbf{w}$ = $\mathbf{\hat{w}}$\\
		}{$final_p = final_p(\text{bigcoeffs})$\\
		  $\mathbf{w} = \text{zeros}(Q,1)$\\	      $\mathbf{w}$($final_p$) = $\mathbf{\hat{w}}(\text{bigcoeffs})$\\
		}
		$\,$ \\
		\% Calculate error and threshold\\
		$e = \|\mathbf{\Theta} ^{\text{test}}\mathbf{w}-\mathbf{g}^{\text{test}}\|^2_2+\eta\|\mathbf{w}\|_0$\\
		\% Adapt threshold\\ 
		\eIf{$e \le e_{\text{best}}$}{ \% Error is decreasing. Increase threshold\\
			$e_{\text{best}} =e$ \\
			$\mathbf{w}_{\text{best}} =\mathbf{w}$ \\
			$tol =tol+d_{\text{tol}}$ \\
		}{
			\% Tolerance too high\\
			 $\mathbf{\Theta}^{\text{train}} = \mathbf{\Theta}^{\text{train}}_{\text{old}}$\\
			$tol$ = max([0,\,$tol-2d_{\text{tol}}$]) \\
			\% Change threshold stepping\\
		$d_{\text{tol}}=\frac{2d_{\text{tol}}}{n_{\text{iters}}-i}$ \\
			$tol=tol+d_{\text{tol}}$ \\
		}
		}
	return $\mathbf{w}_{\mathrm{best}}$\\
	$\,$ \\
	\label{alg1}
	\caption{Pseudocode for automatic threshold sparse Bayesian learning (ATSBL), which combines BCSL~\cite{babacan2009bayesian} and TrainSTRidge~\cite{rudy2017data} to achieve parameter-free inference of highly sparse solutions to inverse problems.}
\end{algorithm}
\subsection{Quality score for identified governing equations}
The error of the inference procedure can be directly quantified by comparison of the results with a known set of original differential equations in test cases. For this purpose, we define the deviation of identified coefficient (DIC) as
 \begin{equation}
\text{DIC}=\frac{1}{K}\sum_{\{i|(c_i\neq0 \,\lor\, c'_i\neq0)\}}\frac{\|c_i-c'_i\|_2}{\text{max}(\|c_i\|_2,\;\|c'_i\|_2)},
\label{eq:5-1b}
\end{equation}
where every $c_i$ is a coefficient of one term in the identified equation and $c'_i$ is the related coefficient in the original equation that was used to generate the test data. Here, at least one of the coefficients in each pair $\{c_i, c'_i\}$ is required to be non-zero and the sum only runs over these coefficients. $K$ represents the number of {these} coefficients. The DIC lies in the range $[0,\infty]$ where $0$ indicates a perfectly identified equation. 

\section{Results} 
\subsection{Inference of SDEs from noisy trajectories}
\label{sec_infSDE1}
\begin{figure*}[!h]
	\centering
	\includegraphics[width=1\linewidth]{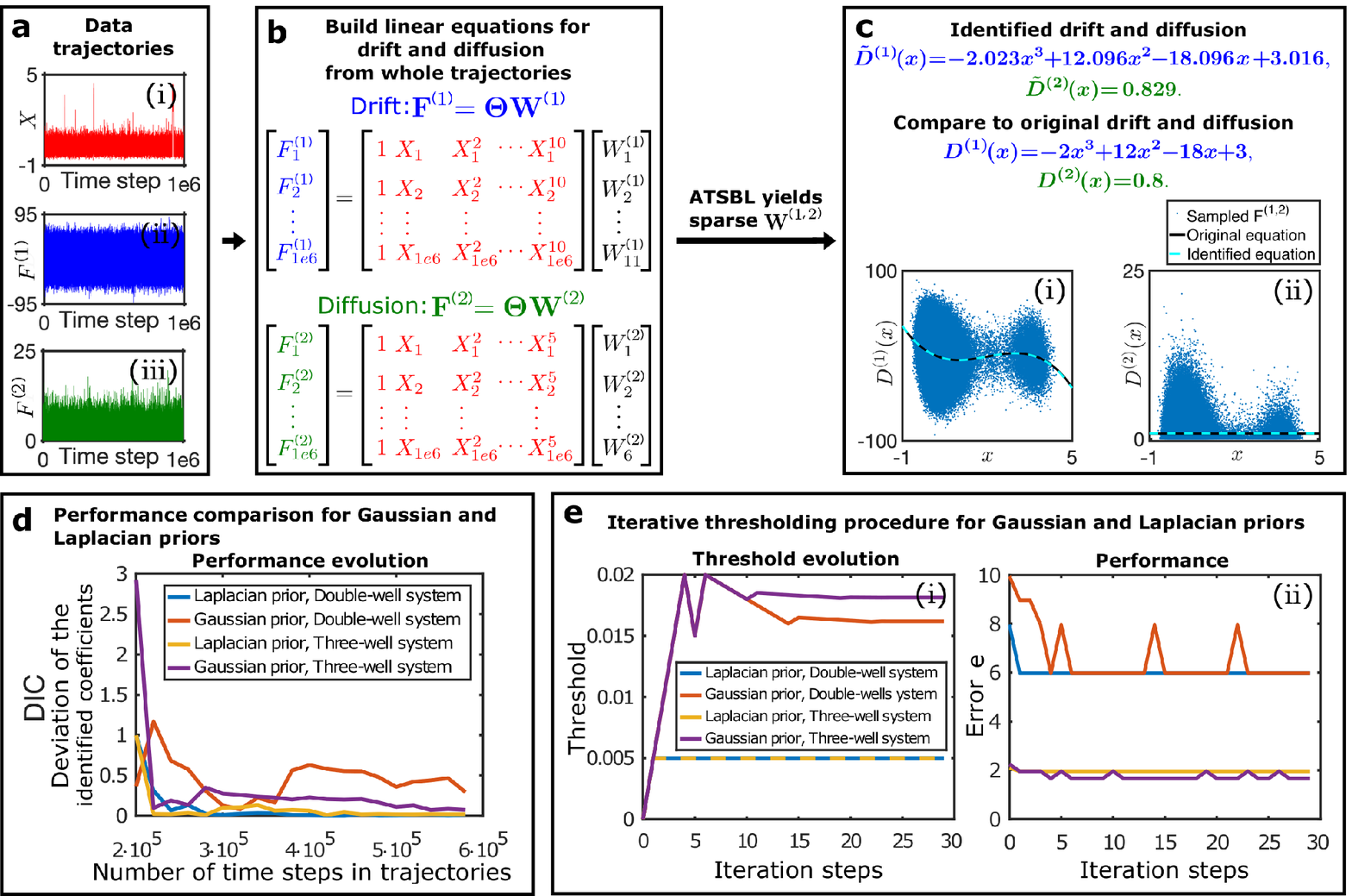}
	\caption{Data-driven discovery of a one-dimensional SDE with automatic threshold sparse Bayesian learning (ATSBL). (a-i)~Trajectory of a particle undergoing overdamped diffusive motion in a double-well potential ($10^6$ time steps). (a-ii,a-II)~Values of the $\mathbf{F}^{(1)}$ and $\mathbf{F}^{(2)}$ generated with discrete differences from the same trajectory.
 (b)~The library matrix $\Theta$ is constructed by evaluating a given set of functions for all values of the trajectory. Thereby, one obtains linear equation systems that relate the known sequences $\mathbf{F}^{(1,2)}$ to unknown, sparsely populated coefficient vectors $\mathbf{W}^{(1,2)}$. The determination of the non-zero entries of $\mathbf{W}^{(1,2)}$ yields a set of library functions that together describe the drift and diffusion coefficients $D^{(1)}$ and $D^{(2)}$. (c)~Exemplary results of the inference procedure. Despite the large noise amplitude, accurate predictions can be made directly from the trajectory data. (d)~Comparison of the use of a Laplacian and Gaussian prior distribution in the inference procedure. The deviation of the identified coefficient (DIC) for the drift coefficient is plotted against the number of data points used for training. The Laplace prior in ATSBL decreases the error and reduces the required sample size. (e)~Convergence rate of the thresholding procedure for Laplacian and Gaussian prior distributions. (e-i)~Laplace priors result in fast threshold convergence.
  (e-ii)~The error $e$ defined in Eq.~(\ref{eq:5-4}) decreases during the iterations. Errors achieved with Gaussian- and Laplacian priors are comparable.
  }
	\label{fig1}
\end{figure*}

We now illustrate the data-driven identification of SDEs by the example of overdamped Brownian motion of a particle inside a one-dimensional double-well potential with coordinate $x$. The drift and diffusion coefficients of this system are given by
\begin{subequations}
\begin{align}
&D^{(1)}(x)=-2x^3+12x^2-18x+3,\\
&D^{(2)}(x)= 0.8.
\end{align}
\end{subequations}
The trajectory data that is to be used for inferring the governing equation is generated by integrating the Langevin equation with the Euler-Maruyama method. A trajectory $\mathbf{X}$ is shown in Fig.~(\ref{fig1}-a-i) ($10^6$ time steps). The trajectories  $\mathbf{F}^{(1)}$ and $\mathbf{F}^{(2)}$ are shown in Fig.~(\ref{fig1}-a-ii, iii). To visualize the $x$- dependence of the estimator for the drift coefficient, we plot $\mathbf{F}^{(1)}$ against $\mathbf{X}$, see Fig.~(\ref{fig1}-c-i). Similarly, $\mathbf{F}^{(2)}$is plotted against  $\mathbf{X}$ as estimator of the diffusion coefficient $D^{(2)}(x)$ in Fig.~(\ref{fig1}-c-ii). 
Both plots exhibit large fluctuations around the true drift and diffusion coefficients and the resulting averages are clearly prone to errors, particularly at the boundaries of the sampled domain.

Using the trajectory data, we next construct a library consisting of $11$ terms for the drift coefficient and $6$ terms for the diffusion coefficient as illustrated in Fig.~(\ref{fig1}-b). Then, we employ ATSBL to identify $\mathbf{W}^{(1)}$ and $\mathbf{W}^{(1)}$ directly from the trajectory without binning. 
The identified $x$-dependent functions for the drift and diffusion coefficients are plotted in Fig.~(\ref{fig1}-c-i, ii). They agree well with the original functions used for creating the data. The identified equations with estimated uncertainties are shown in Fig.~(\ref{fig1}-c).

The main distinction of ATSBL as compared to established inference techniques is the assumption of a Laplacian distribution for the prior of the library coefficients. The more direct, albeit theoretically less sparsity-promoting procedure is to employ a Gaussian prior, corresponding to a Ridge regression with fixed regularization parameter, for inference of the solution vector $\mathbf{w}$ in Eq.~(\ref{eq:2-2}) prior to automatic thresholding, as done, e.g., in Ref.~\cite{rudy2017data}. To compare the performance of these two approaches for inference of SDEs, we evaluate the deviation of the identified coefficients, DIC, as a function of the number of data points used for inference. The results shown in Fig.~(\ref{fig1}-d) indicate that the Laplacian prior is preferable over the  Gaussian prior since it requires less data and results in a smaller DIC. To further establish the robustness of ATSBL, we consider the convergence of the iterative thresholding procedure for each of the two prior distributions. The result shown in Fig.~(\ref{fig1}-e-i, ii) demonstrate a better convergence achieved in the case of the Laplacian prior. For both, Gaussian and Laplacian prior distributions, the threshold and the error oscillate during the iteration process, which is due to the adaptive step size during the thresholding.

\subsection{Inference of SDEs with time-dependent drift coefficient}
In the previous section, an example is provided of how the KM coefficients can be obtained by performing a regression directly with the trajectory data. The direct use of the trajectory data becomes particularly important for the treatment of the more complex situation of a time-varying force. In such a situation, the probability distributions change over time and a histogram-based approximation of the dynamic distributions can be technically challenging and requires the availability of many sample trajectories for the same conditions. In order to explore the validity of our approach in this situation, we consider the example of a particle diffusing within a time-dependent one-dimensional potential. The drift and diffusion coefficients of this system are given by
\begin{subequations}
\begin{align}
&D^{(1)}(x)=\left[a_{0}+1-\cos(\omega t)\right]x-x^{3},\\
&D^{(2)}(x)= 0.8,
\end{align}
\end{subequations}
with $a_{0}=5 \cdot 10^{-3}$. The potential has a double-well shape, where the positions of the two minima vary in time. The two minima start at separate positions and merge periodically into one minimum before separating again to reach the initial positions. Each time the potential wells come close to each other, the transition probability becomes large and the particle is likely to change from one well to another. This gives rise to stochastic oscillations between the two potential wells. To test whether the underlying equations can be inferred with a library constructed from a single trajectory, we assume that the frequency $\omega$ at which the potential changes is known. Inference of this frequency from the data is in principle also possible, but requires excessive computational power since high-order terms with explicit time-dependence must be accounted for in the library. We construct a library consisting of a time-dependent and a time-independent part. The first half of the library is simply a polynomial basis, the second half corresponds to the polynomial basis multiplied with a $\cos(\omega t)$ factor. The results of the inference procedure are shown in Fig.~(\ref{fig2}-a). The inferred equation is in agreement with the correct equation. For illustration, we plot snapshots of the drift and diffusion for the original equations and the inferred equations in  Fig.~(\ref{fig2}-a-ii, iii). Note that the inference procedure for first-order SDEs shows a remarkable performance even though only one sample trajectory is used for inference. 
\subsection{Data binning for inference of SDEs from short trajectories}
\begin{figure*}[!h]
	\centering
	\includegraphics[width=1\linewidth]{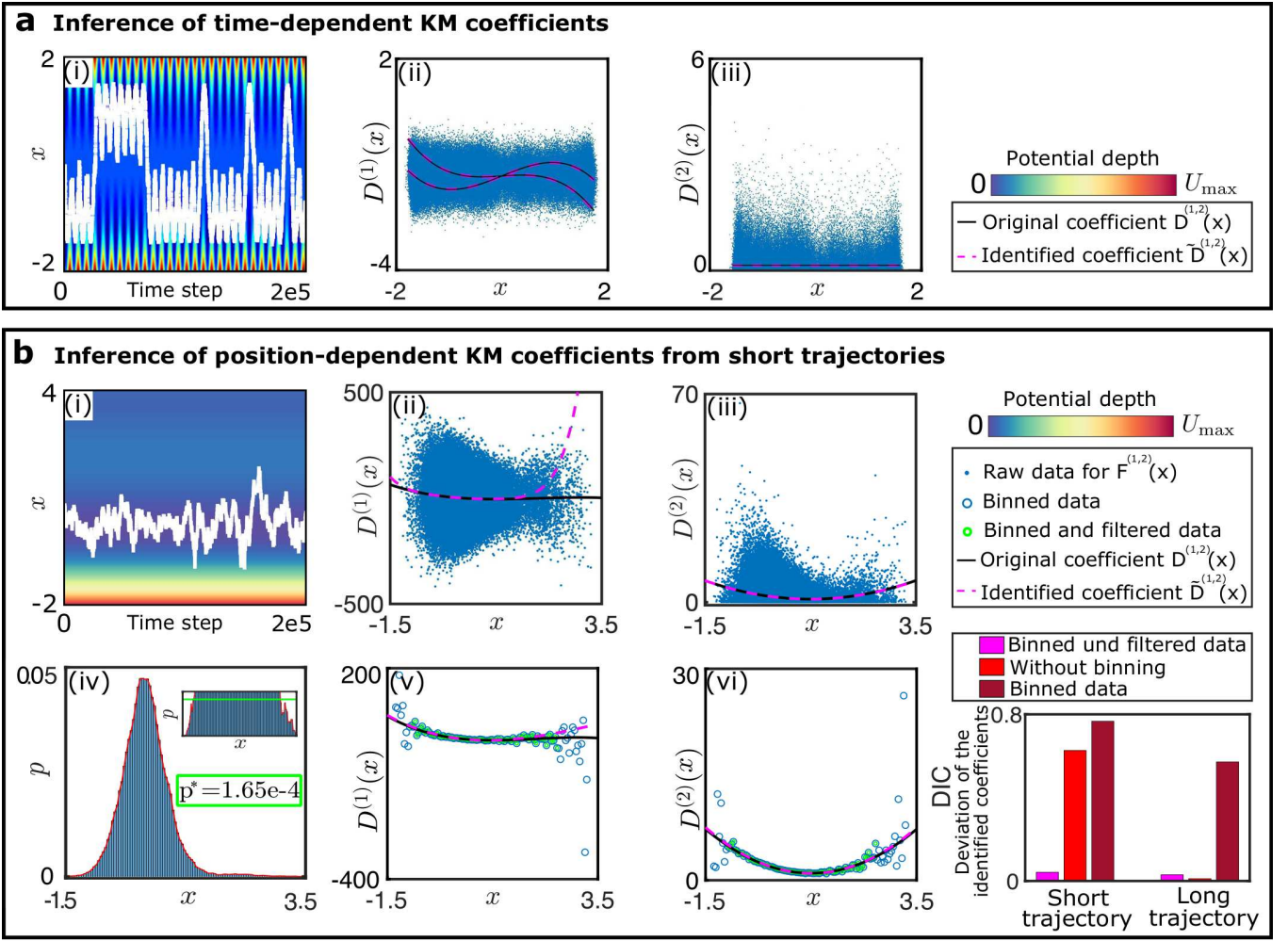}
	\caption{Inference of KM coefficients for one-dimensional SDEs. 
	(a)~System with a time-dependent force field. 
	(a-i)~Trajectory of an overdamped motion in a time-varying double-well potential. 
	(a-ii, iii)~Using an appropriate function library, the functional forms of the KM coefficients can be faithfully reconstructed. Blue dots are values of $F^{(1,2)}$ estimated from the trajectory.
	(b)~Advantage of data binning for analysis of short trajectories. 
	(b-i)~Trajectory resulting from overdamped motion in a double-well potential with space-dependent diffusion coefficient.  
	(b-ii, iii)~Inferred x-dependence of the KM coefficients for short trajectories ($2 \cdot 10^5$ time steps). The unpopulated regions in phase space are characterized by a high uncertainty of inference and therefore lead to large deviations in the coefficients.
	(b-iv)~Histogram of particle positions for the trajectory shown in (i). (b-iv, v, vi)~Binned distributions can be used to infer the KM coefficients, but large errors occur in regions that are not well-sampled. Inference errors due to incomplete phase-space sampling for short trajectories can be accounted for by excluding the data below a probability threshold, corresponding to large uncertainty. (b-vii)~Performance of the inference with data binning and without data binning for short and long trajectories ($2 \cdot 10^5$ and $2 \cdot 10^7$ time steps, respectively). The shown DIC is the average of the DICs for $D^{(1)}$ and $D^{(2)}$. For long trajectories, data binning does not reduce the error.}
	\label{fig2}
\end{figure*}

An inference method that relies on direct use of sample trajectories for a regression can become unreliable when confronted with short trajectories in an inhomogeneous force field. In such a situation, we find that it is more appropriate to employ data binning. We illustrate this procedure with Brownian motion of a particle in a one-dimensional double-well potential where the diffusion coefficient depends on space. The drift and diffusion coefficients of the model are given by  
\begin{subequations}
\begin{align}
&D^{(1)}(x)=-2x^3+12x^2-18x+3,\\
&D^{(2)}(x)= x^2-2x+2.
\end{align} 
\end{subequations}
We first consider short trajectories that have $2\cdot10^5$ time steps, exemplified by the plot in Fig.~(\ref{fig2}-b-i). The raw data and the binned data are shown in
Fig.~(\ref{fig2}b-ii, iii) and Fig.~(\ref{fig2}b-v, vi), respectively. For this example, $200$ data bins are employed. The distributions approximated by the binned data clearly deviate from the known functions $D^{(1)}(x)$ and $D^{(2)}(x)$ in undersampled regions. Therefore, the binned data is filtered to remove data points with high uncertainty. This filtering is done as described in the Methods section by discarding bins below a probability threshold $p^*$ that is determined by entropy maximization~\cite{el2011new}, see Fig.~(\ref{fig2}-b-iv). To assess if binning and filtering is also beneficial for inference from long trajectories, we also use data from trajectories with $2\cdot10^7$ time steps. Figure~(\ref{fig2}-b-vii) shows the errors of the identified coefficients.

For short trajectories, binning is advantageous in combination with a filtering procedure to suppress data with high uncertainty. The reason for this result can be understood from inspection of Fig.~(\ref{fig2}b-v, vi), where the inferred functions match the correct functions only in the most populated regions of phase space. Thus, the exclusion of data points with high uncertainty prevents overfitting and improves the inference of the underlying dynamical equations if the trajectory is not long enough to allow a sufficient sampling of the whole phase space. Conversely, data binning with or without filtering is disadvantageous for the analysis of long trajectories that sample the whole phase space, see Fig.~(\ref{fig2}-b-vii).

\subsection{Active sampling improves the identification of SDEs}
We have so far restricted our attention to the extraction of estimates from data that was generated prior to the analysis, e.g., in experiments. Thereby, we have assumed that the size of the data set is large enough to allow some form of inference of the governing equations. In a different scenario one might have the ability to perturb the studied system, either in a computer simulation or in an experimental setup, while simultaneously recording the data. Then, one may enhance the sampling efficiency by means of an appropriately designed perturbation that is applied to the system. Generally, this methodology is expected to be useful whenever the system exhibits an energy landscape with multiple local minima that can trap the trajectory for long times. We describe an adaptive control method where the inference of the dynamical equations together with a simultaneous perturbation of the system recursively results in a global exploration of the phase space to provide sufficient sampling everywhere.

Since the probability distribution tends to be peaked around local energy minima, the dynamical equations can be estimated locally near these minima. To take advantage of this local estimation while iteratively extending the sampled region, we re-sample repeatedly while applying in each sampling round a control force that is opposite to a force from the system that is estimated locally from previous rounds. The difficulty with a straight-forward application of this method is that the control force can admit large deviations away from the initial estimation region. This effect produces large errors, slows down convergence, and may even lead to divergence problems. We overcome this problem by weighting the control force with a Gaussian distribution such that the control force vanishes away from the current estimation region. Such a control force expels the trajectory from the local minimum where the estimation has been performed and the trajectory eventually reaches another local minimum. 

\begin{figure*}[!h]
	\centering
	\includegraphics[width=1\linewidth]{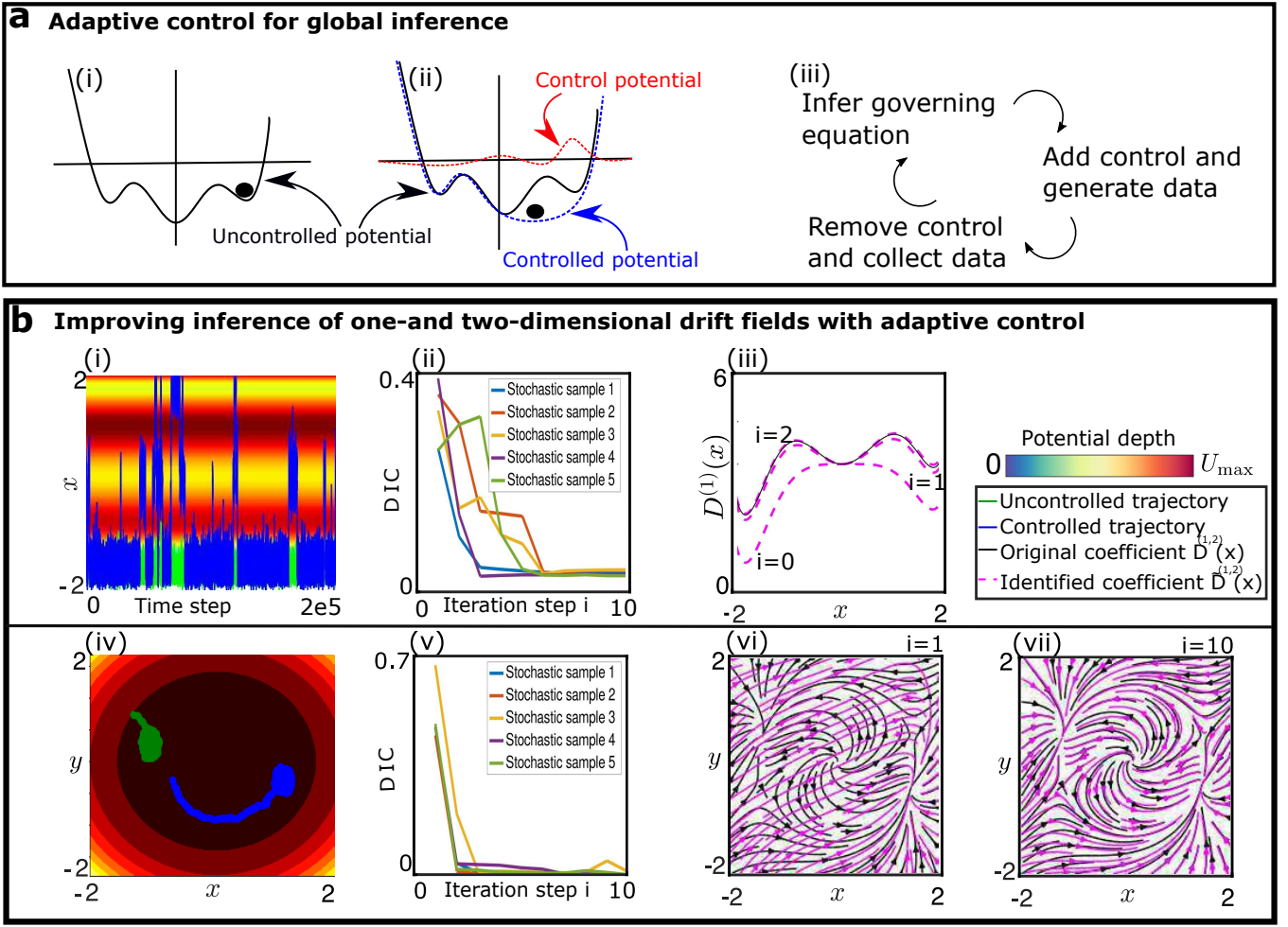}
	\caption{Active learning with automatic iterative sampling optimization (AISO). (a)~Schematic presentation of automatic iterative sampling optimization for the case of Brownian diffusion (a-i). Initially, the particle is trapped in a local energetic minimum and the functional form of the potential can therefore only be inferred locally. (a-ii)~After the first iteration step, the potential hypersurface near the estimated minimum is flattened and the particle  can thus explore other regions of phase space. The same procedure is repeated iteratively and the control is always applied at the minimum estimated during the previous iteration. (a-iii)~Schematic representation of the main feedback control loop. (b-i)~Trajectory of a particle undergoing Brownian motion in a one-dimensional three-well potential. The green curve shows a trapped trajectory while the blue curve shows a trajectory in the presence of control forces. (b-ii)~The deviation of the inferred coefficients (DIC) decreases during the iterations. (b-iii)~The identified drift field converges to the correct function during the iteration.
(b-iv)~Trajectory of a particle undergoing diffusion in a two-dimensional force field. The green curve exemplifies a trapped trajectory for plain sampling. The blue curve shows an example of a trajectory in the presence of control forces. The color of the background only represents part of the force field, namely a Mexican hat potential $V(x, y) = -(x^2 + y^2)/2 + (x^2 + y^2)^2/4$ that generates radial forces. (b-v)~Evolution of the of the DIC during the iterations. (b-vi)~Streamlines of the identified drift field (pink) and streamlines of the correct drift field (black) after the first iteration step.  (b-vii)~Streamlines of identified drift field (pink) and streamlines of the correct drift field (black) after the tenth iteration step. The identified force field at the end of the iteration closely matches the original one.}
	\label{fig3}
\end{figure*}

The method, which we call automatic iterative sampling optimization (AISO), is illustrated in Fig.~(\ref{fig3}-a, i-iii) and the pseudocode is provided in Algorithm~\ref{alg2}. At each iteration, the underlying dynamical equations are estimated from the data accumulated during all previous iterations. The negative of the inferred drift term is employed locally as control force. The center and width of the Gaussian weight of the control force is calculated only from the mean and standard deviation of the trajectory of the previous step.
Thus, we define our control force acting on the component $\ell$ as
\begin{align}
c^i_{\ell}(\{X_{\ell'}(t)\}) = -\boldsymbol{\Theta}_{\ell}(\{X_{\ell'}(t)\})\cdot\mathbf{w}^i\,\exp{\left[-\dfrac{\parallel X_{\ell}(t)-\mu_{\ell}^{i} \parallel^2}{\zeta_{\ell}^{i}}\right]} 
\label{eq:controlforce}
\end{align} 
where the index $i$ indicates that values are to be taken at the iteration number $i$; $\mu_{\ell}^{i}$ and $\zeta_{\ell}^{i}$ stand for the mean and variance of the trajectory extracted from the step $i$ in each iteration.
After a sufficient number of iterations, the data points accumulated from all iterations are combined and the equation of motion is extracted from the accumulated data. This procedure is repeated for a predefined number of iteration steps. For the examples presented in the following, the iteration step number has been fixed to $N=10$, since convergence has been achieved within less than $10$ steps in these cases.

For a first demonstration of our method, we employ a three-well potential $U(x)=x^6-6x^4+0.5x^3+8x^2$ with a constant diffusion coefficient for simulating the trajectory of a particle in one dimension. The drift and diffusion coefficients are
\begin{subequations}
\begin{align}
&D^{(1)}(x)=-\frac{d U}{d x}=-6x^5+24x^3-1.5x^2-16x,\\
&D^{(2)}(x)= 1.
\end{align} 
\end{subequations}
Next, we also consider a two-dimensional drift field, consisting of a radially symmetric component and a shear component in the $x,y$-plane. The drift and diffusion coefficients are given by
\begin{subequations}
\begin{align}
&D_{x}^{(1)}(x,y)=x(1-x^{2}-y^{2})+y(x^{2}-y^{2}-b)\\
&D_{y}^{(1)}(x,y)=y(1-x^{2}-y^{2})+x(x^{2}-y^{2}-b), \\
&D_{x}^{(2)}(x,y)=D_{y}^{(2)}(x,y)= 1.
\end{align} 
\end{subequations}
Using these driving forces, we simulate trajectories with $10^5$ time steps with one time step being $\Delta t=5\cdot10^{-3}$. Parts of the trajectories on the potential maps are shown in Fig.~(\ref{fig3}-b-i, iv). Results for the intermediate iteration steps are shown together with the drift field in Fig.~(\ref{fig3}-b-iii, vi). As the algorithm proceeds through more iterations, the coefficients of the control potential approach the coefficients of the correct drift field, and the expulsion from each local minimum becomes more efficient, Fig.~(\ref{fig3}-b-iii, vii). This results in an enhancement of rare events where the particle crosses the saddle points, as illustrated in Fig.~(\ref{fig3}-b-i, iv) by the controlled and uncontrolled trajectories. The error is quantified by calculating the coefficients $\tilde{D}^{(1)}(x)$ and $\tilde{D}^{(2)}(x)$ in each iteration. The DIC reduces  from $1$ to nearly $0.01$ during the iterations, see Fig.~(\ref{fig3}-b-ii, v). Thus, the terms of the identified equations approach those given in the original equations.

\begin{algorithm}
	\SetAlgoLined
	\SetKwFor{For}{for (}{) $\lbrace$}{$\rbrace$}
    \SetKwInput{KwFunction}{Function}
	\KwFunction{AISO()}
	$\mathbf{c}^0 = \mathbf{0}$; \space\space\%Initialize control force to zero \\
	N = 10; \%Fix number of iteration steps \\
	i = 0; \space\space\space\%Initialize iterator to zero \\
	\While{$i<N$}{
		i= i+1;\\
		\%For all degrees of freedom $\ell,\ell' \in \{1\ldots M\}$ \\
		\%Add control force and generate data\\
		$\frac{dX_{\ell}(t)}{dt}= g_{\ell}(\mathbf{X}(t),t)+ c_{\ell}^{i-1} +  
		\sum_{\ell'}h_{\ell,\ell'}(\mathbf{X}(t),t)dW_{\ell'}(t)$ \\	   
		\%Subtract control force and collect data\\
		$\frac{dX_{\ell}(t)}{dt} \leftarrow \frac{dX_{\ell}(t)}{dt} - c_{\ell}^{i-1}$ \\
		\%Concatenate data\\
		$\Theta \leftarrow (\Theta, \Theta(X_{\ell}(t))$ \\ 
		$\mathbf{g} \leftarrow (\mathbf{g}, \frac{dX_{\ell}(t)}{dt})$ \\
		\%Estimate libary coefficients with ATSBL \\
		$\mathbf{w}$ = ATSBL($\mathbf{\Theta}$, $\mathbf{g}$, $d_{\text{tol}}$, $n_{\text{iters}}$) 
		}
		return $\mathbf{w}$
		\, 
	\label{alg2}
	\caption{Pseudocode for identification of SDEs with automatic iterative sampling optimization (AISO)}
\end{algorithm}

\subsection{Identification of ordinary and partial differential equations}
\begin{figure*}[!h]
	\centering
	\includegraphics[width=1\linewidth]{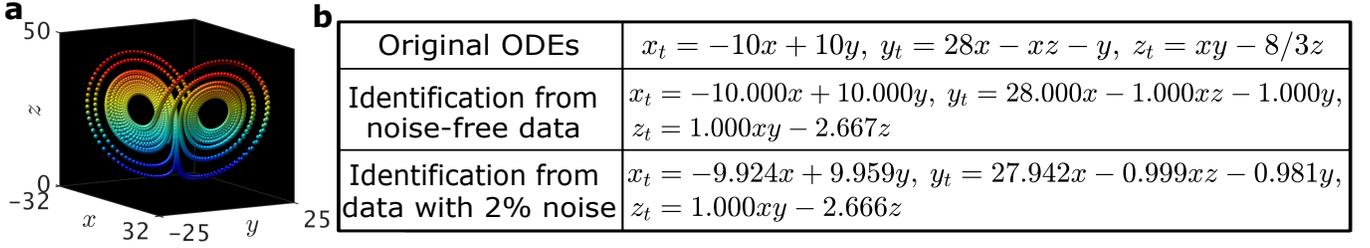}
	\caption{Example for data-driven discovery of ODEs with ATSBL. (a)~Plot of a numerically integrated trajectory for $t\in [0, 25]$ with a time step of $\Delta t=2 \cdot 10^{-4}$ and an initial condition as $[x_0,\;y_0,\;z_0]=[-8,\;8 ,\;27]$. (b)~The table shows the original ODEs, i.e., the Lorenz system, and the identified ODEs from noise-free data and data with 2\% Gaussian noise.}
	\label{fig4}
\end{figure*}

It is next shown that the sparse inference scheme based on Laplace priors that is implemented with ATSBL can also be used for data-driven discovery of ordinary and partial differential equations. The identification of ODEs from trajectory data is demonstrated with a Lorenz system, which is a paradigm for chaotic behaviour~\cite{lorenz1963deterministic}. The Lorenz equations are given by 
\begin{subequations}
\begin{align}
x_t&=a(y-x),\\
y_t&=x(b-z)-y,\\
z_t&=xy-cz,
\end{align}
\label{eqn:4-5}%
\end{subequations}
where the subcript $t$ denotes a time derivative and the parameters are fixed as $a=10$, $b=28$, and $c=3/8$. We numerically integrate these equations to obtain a trajectory as shown in Fig.~(\ref{fig4}-a). The chaotic system involves two attractors. For data-driven system identification, we utilize three identical libraries $\mathbf{\Theta}$ for each of the variables, $x$, $y$ and $z$. $\mathbf{\Theta}$ is constructed from the simulated trajectory and includes 56 terms containing up to fourth powers of all variables. Time-derivatives are calculated using fourth-order central-difference approximation. The general ODEs constructed from the library as in Eq.~(\ref{eq:5-40}) are represented by three linear equation systems. {The estimated equations resulting from an application of the inference procedure to noise-free data have small errors that are in magnitude comparable to the time step $\Delta t = 2 \cdot 10^{-4}$,} see Fig.~(\ref{fig4}-b). The same inference procedure is then repeated for a trajectory with additive Gaussian noise. The standard deviation of the noise in each coordinate is chosen to be 2~\% of the standard deviation of the noise-free data in the same coordinate. For this case, the ODEs identified with ATSBL still contain all the correct terms and the errors in the inferred system parameters are in the percent range, see Fig.~(\ref{fig4}-b). 

\begin{figure*}[!h]
	\centering
	\includegraphics[width=1\linewidth]{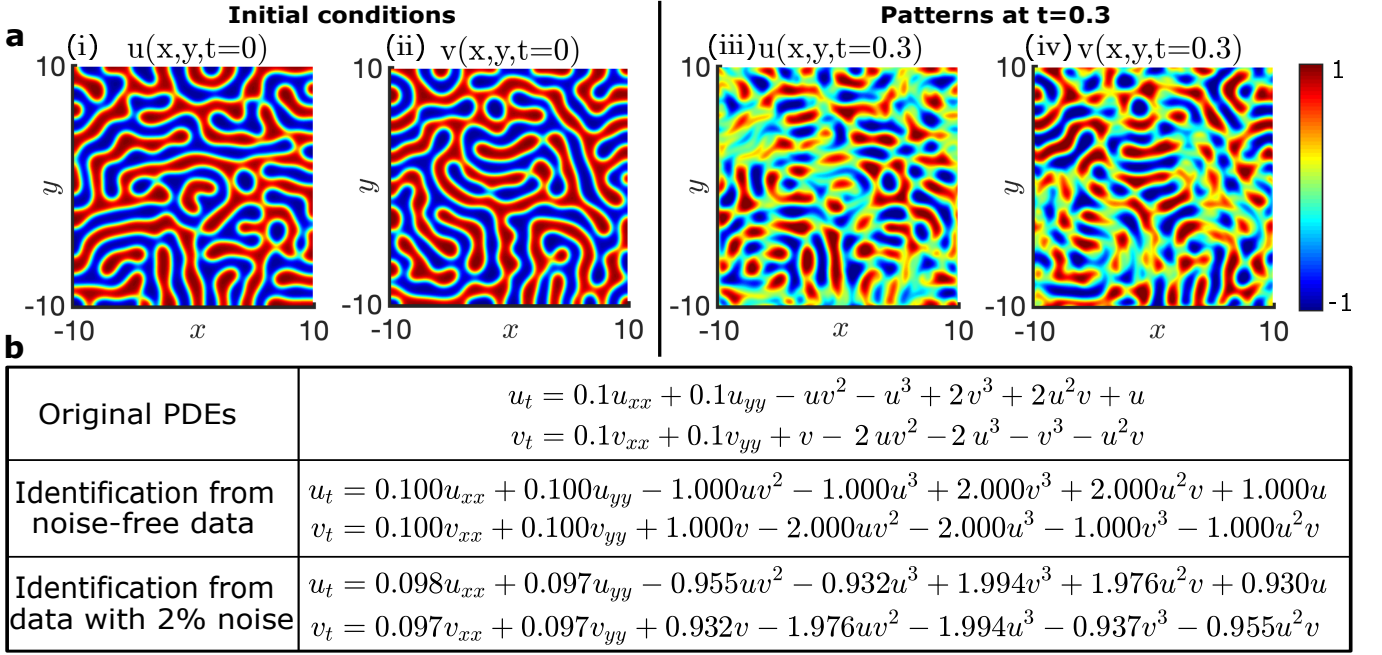}
	\caption{Demonstration of data-driven discovery of PDEs with ATSBL using a reaction-diffusion system. (a-i, ii)~Snapshots of the initial conditions for the variables $u$ and $v$, respectively. (a-iii, iv)~ $u$ and $v$ at time $t=0.3$.  
	(b)~The table shows the original PDEs for the reaction-diffusion system and the identified PDEs for noise-free data and data with 2\% Gaussian noise. Inference is conducted with a library containing $35$ terms. The numerical calculations are done with a time step $\Delta t=0.0034$ in the time interval $t=[0,0.6]$. The space domain has size $20\times20$ and is covered with a $256\times 256$ grid with periodic boundary conditions.
	}
	\label{fig5}
\end{figure*}

Finally, we demonstrate data-driven discovery of PDEs with ATSBL. Reaction-diffusion equations have attracted interest as prototypic models for pattern formation in biochemical systems, where constituents are locally transformed into each other through chemical reactions and transported in space by diffusion. Here, we consider the popular $\lambda-\omega$ system, given by
\begin{subequations}
	\begin{align*}
	&u_t=D_u\nabla^2u+\lambda(A)u-\omega(A)v,\\
	&v_t=D_v\nabla^2v+\omega(A)u+\lambda(A)v,\\
	&A=u^2+v^2,\;\;\omega=-\beta A^2,\;\; \lambda=1-A^2,
	\end{align*}
	\label{eq:5-9}%
\end{subequations}
where $\beta$ is equal to 2. A two-dimensional, planar, rectangular area with periodic boundary conditions is considered. The initial values of $u$ and $v$ are shown in Figs.~(\ref{fig5}-a-i, ii). The reaction-diffusion equations are solved numerically by using a spectral method. Snapshots of $u$ and $v$ are shown in Figs.~(\ref{fig5}-a-iii, iv). For inference of the governing PDEs, a library matrix $\mathbf{\Theta}$ is constructed containing 35 terms each for $u_t$ and $v_t$. Then, using ATSBL, the reaction-diffusion equations are inferred from the simulated data, as illustrated in Fig.~\ref{fig5}(b). For noise-free data, the identified equations deviate from the original equations only at the fourth decimal place and this error is due to discretization.
However, if $u$ and $v$ are corrupted with additive noise, identification of the correct PDEs becomes challenging~\cite{rudy2017data}. In Ref.~\cite{rudy2017data}, it has therefore been suggested to include a denoising step prior to the inference step. Accordingly, we employ a curvelet denoising method~\cite{peyre2011numerical}, which permits reconstruction of the reaction-diffusion equations from data with 2\% noise with ATSBL, as illustrated in Fig.~\ref{fig5}(b).  

\section{Summary and Outlook}
Data-driven, automatic discovery of governing equations has become a viable tool for studying complex systems if first-principle derivations are intractable, e.g., for biological systems or epidemiological data. The aim is here generally to construct an analytical model that characterizes the observed dynamics and extends to parameter- and phase space regions that are hard to access experimentally.

Our main contribution is an inference method that makes use of Laplacian prior distributions in a Bayesian framework to find a minimal set of governing equations without the need for user input. We establish the validity of this approach and compare it to other methods. Regarding data-driven discovery of Langevin-type SDEs, we show that the proposed sparse method converges faster than other methods based on ridge regression. {Maximum likelihood methods for the estimation of parameters in SDEs are not considered here, see Ref.~\cite{bishwal2007parameter} for an introduction to those methods.} For the studied Langevin SDEs, we find that a binning of the trajectory data for inference of the drift and diffusion coefficients is only advantageous if the phase space is sampled sparsely. In that case, the error of the inference procedure can vary strongly in phase space since the relative uncertainties of the probabilities vary. A filtering procedure consisting of the exclusion of data with high uncertainty results in a significantly improved inference accuracy.

Next, we investigate how active-learning procedures can be useful in situations where the inference of globally valid equations becomes difficult because most trajectories are trapped in local potential minima. {This problem can be solved} with well-established umbrella-sampling methods where quadratic bias potentials are employed to reduce the energetic barriers in the original potential landscape~\cite{torrie1977nonphysical,bussi2020using}. However, {an appropriate parameterization of such bias potentials can be challenging.} For example, if the additional potentials are intended to smoothen an unknown, rough potential landscape. Instead, we employ data-driven identification of governing equations for calculating time-dependent external perturbations that force the trajectory to explore the full phase space. The main feature of our method is that the parameters that determine the control force correspond to the parameters that define the potential landscape. {The combination of iterative inference with system perturbations can significantly improve the speed and accuracy of the overall inference procedure.} We therefore hope that the suggested active learning procedure will extend the applicability of data-driven methods, in particular in the context of computer simulations.

A central challenge related to the improvement of the library-based methodology for  identification of analytical models is to find automated approaches for tailoring the employed function space to the problem at hand. Recent methodological advances suggest that a possible solution is the integration of physical constraints, such as symmetries, conservation laws, or even thermodynamics, into a generic framework for statistical learning of governing equations~\cite{karniadakis2021physics}. Data-driven identification of analytical models thus has the potential to become a popular tool for closing the gap between non-parametric, empirical modeling and first-principles-based modeling in the coming years. 

\begin{acknowledgments}
Funding by the European Research Council through a starting grant for BS is gratefully acknowledged (BacForce, g.a.No.~852585).
\end{acknowledgments}

\section*{Author contributions}
All authors designed the study, performed the research and wrote the manuscript together.

\section*{Competing interests}
The authors declare no competing interests.\\

\section*{Data availability}
Source code and data can be obtained from the corresponding author upon request.


\newpage

\end{document}